\documentclass[11pt]{article}
\usepackage{mathptmx}
\usepackage{sprinconf}
\usepackage{graphicx}
\usepackage{bm}
\usepackage{amsmath}
\usepackage{amsfonts}
\usepackage{amscd}

 \font\tensym=msbm10
 \font\sevensym=msbm7
 \font\fivesym=msbm5

 \font\tengoth=eufb10
 \font\sevengoth=eufb7
 \font\fivegoth=eufb5

\newfam\symfam
\textfont\symfam=\tensym
\scriptfont\symfam=\sevensym
\scriptscriptfont\symfam=\fivesym

\newfam\gothfam
\textfont\gothfam=\tengoth
\scriptfont\gothfam=\sevengoth
\scriptscriptfont\gothfam=\fivegoth

\def\hs{\hbox to 3mm{}}
\def\hhs{\hbox to 5cm{}}

\def\bs{\bigskip}

\def\JPicScale{0.8}\ifx\JPicScale\undefined\def\JPicScale{1}\fi

\def\N{\mathbb{N}}

\def\HWS{{\cal H}_{WS}}
\def\1H{\mathbf{1}_{\HWS}}
\def\L{\mathbb{L}}
\def\V{\mathbb{V}}

\def\al{\alpha}
\def\be{\beta}
\def\ep{\varepsilon}

\def\om{\omega}

\def\diag{\mathbf{diag}}
\def\ldiag{\mathbf{ldiag}}
\def\DIAG{\mathbf{DIAG}}
\def\LDIAG{\mathbf{LDIAG}}
\def\MQS{\mathbf{MQSym}}

\def\ra{\rightarrow}

\def\ua{\uparrow}

\def\adots{\mathinner{\mkern2mu\raise1pt\hbox{.}
\mkern3mu\raise4pt\hbox{.}\mkern1mu\raise7pt\hbox{.}}}

\def\up#1{\raise 1ex\hbox{\footnotesize#1}}

\newtheorem{expl}{Exemple}[section]

\newtheorem{proposition}[expl]{Proposition}

\begin{document}
\title{A multipurpose Hopf deformation of the Algebra of Feynman-like Diagrams}
\author{G H E Duchamp$^{a}$, A I Solomon$^{c,d}$, P Blasiak$^{b}$, K A Penson$^{c}$ and A Horzela$^{b}$}
\affil{$^a$ Institut Galil\'ee, LIPN, CNRS UMR 7030\\
99 Av. J.-B. Clement, F-93430 Villetaneuse, France\vspace{2mm}\\
$^b$ H. Niewodnicza\'nski Institute of
Nuclear Physics\\
Polish Academy of Sciences\\
ul. Eliasza-Radzikowskiego 152,  PL 31342 Krak\'ow, Poland\vspace{2mm}\\
$^c$ Laboratoire de Physique Th\'eorique de la Mati\`{e}re Condens\'{e}e\\
Universit\'e Pierre et Marie Curie, CNRS UMR 7600\\
Tour 24 - 2i\`{e}me \'et., 4 pl. Jussieu, F 75252 Paris Cedex 05, France\vspace{2mm}\\
$^e$ The Open University, Physics and Astronomy Department\\
Milton Keynes MK7 6AA, United Kingdom}


\beginabstract
We construct a three parameter deformation of the Hopf algebra $\LDIAG$. This new algebra is a true Hopf deformation which  reduces to $\LDIAG$ on one hand and to $\MQS$ on the other, relating $\LDIAG$ to  other Hopf algebras of interest in contemporary physics. Further,  its product law reproduces that of the algebra of polyzeta functions.
\endabstract

\section{Introduction}

The complete journey between the first appearance of a product formula by Bender et al. \cite{BBM} and  their related Feynman-like diagrams to the discovery of a Hopf algebra structure \cite{GOF4}  on the diagrams themselves, goes roughly as follows.\\
Firstly, Bender, Brody, and  Meister \cite{BBM} introduced a special field theory which proved to be particularly rich in combinatorial links and by-products \cite{GOF8} (not to mention the link with vector fields and one-parameter groups \cite{OPM,GOF7}).\\
Secondly, the Feynman-like diagrams of this theory label monomials which combine  naturally in a way compatible with  monomial multiplication and co-addition (i.e. the standard Hopf algebra structure on the space of polynomials). This is the Hopf algebra $\mathbf{DIAG}$ \cite{GOF4}. The (Hopf-)subalgebra of $\mathbf{DIAG}$ generated by the primitive graphs is the Hopf algebra $\mathbf{BELL}$ described in Solomon's talk at this conference \cite{GOF9}.\\
Thirdly, the natural noncommutative pull-back of this algebra, $\LDIAG$, has a basis (the {\em labelled} diagrams) which is
in one-to-one correspondence with that of the {\em Matrix Quasi-Symmetric Functions} \cite{DHT} (the {\em packed matrices} of $\MQS$), but their algebra and co-algebra structures are completely different. In particular,  multiplication in $\MQS$ involves a sort of shifted shuffle with overlappings reminiscent of  Hoffmann's shuffle used in the theory of polyzeta functions \cite{Ca}. The superpositions and overlappings involved there are not present in  (non-deformed) $\LDIAG$ and, moreover, the coproduct of $\LDIAG$ is co-commutative while that of $\MQS$ is not.

The aim of this paper is to announce the existence of a Hopf algebra deformation which connects $\LDIAG$ to other Hopf algebras relevant to physics (Connes-Kreimer, Connes-Moscovici, Brouder-Frabetti, see \cite{Fo2}) and other fields (noncommutative symmetric functions, Euler-Zagier sums).\\
{\sc Acknowledgements} : The authors would like to thank Jim Stasheff, Christophe Tollu and Lo\"\i c Foissy for fruitful interactions.

\section{Labelled Diagrams and Diagrams}

Product formula involves a summation over all diagrams of a certain type \cite{GOF9} a labelled version of which is described below.\\ 
Labelled diagrams can be identified with their weight functions which are mappings $\om : \N^+\times\N^+\ra \N$ such that the supporting subgraph 
\begin{equation}
\Gamma_\om=\{(i,j)\in \N^+\times\N^+\ |\ w(i,j)\not=0\} 
\end{equation}
has specific projections i.e. $pr_1(\Gamma_\om)=[1..p];\ pr_2(\Gamma_\om)=[1..q]$ for some $p,q\in \N$ (notice that when one of $p,q$ is zero so too is the other and the diagram is empty).\\
These graphs are represented by labelled diagrams as follows

\begin{center}
\includegraphics[scale=0.2]{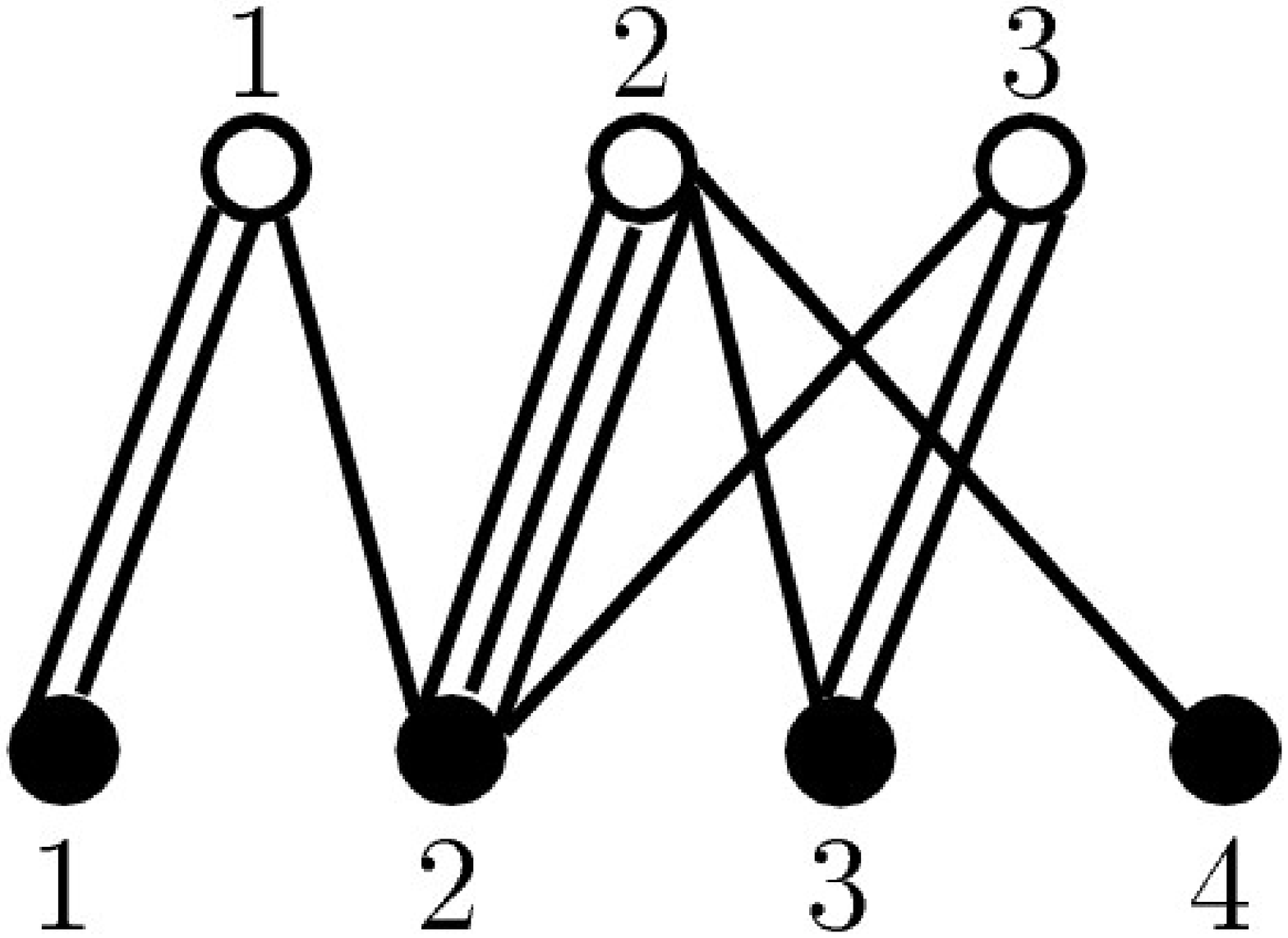}
\end{center}

The labelled diagrams form the set $\ldiag$ and prescribe monomials through the formula $\L^{\al(d)}\V^{\be(d)}$ where $\al(d)$ (resp. $\be(d)$) is the ``white spot type'' (resp. the ``black spot type'') i.e. the multi-index $(\al_i)_{i\in \N^+}$ (resp. $(\be_i)_{i\in \N^+}$) such that $\al_i$ (resp. $\be_i$) is the number of white spots (resp. black spots) of degree $i$ ($i$ lines connected to the spot).\\
There is  a (graphically) natural multiplicative structure on $\ldiag$ such that the arrow 
\begin{equation}
m_{(\L,\V)} :    d \mapsto \L^{\al(d)}\V^{\be(d)}
\end{equation}
is a morphism.\\
It is clear  that one can permute black spots, or white spots, of $d$ without changing the monomial $\L^{\al(d)}\V^{\be(d)}$. The classes of (labelled) diagrams up to this equivalence (permutations of white - or black - spots among themselves) are naturally represented by unlabelled diagrams and will be denoted $\diag$ (including the empty one). 

\begin{center}
\includegraphics[scale=0.4]{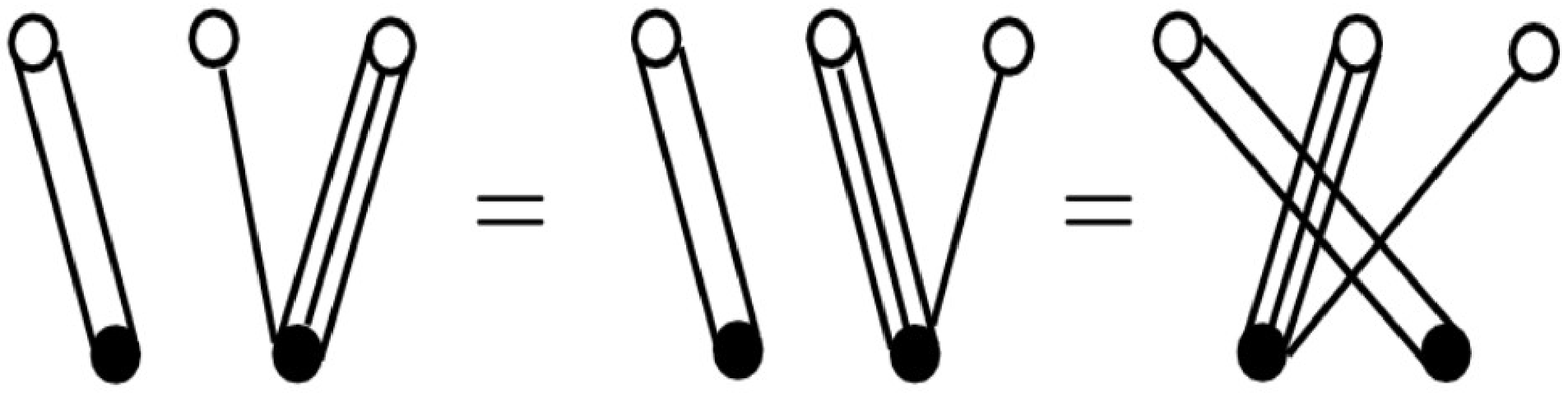}
\end{center}

For both types of diagram the product consists of concatenating the diagrams i.e. placing $d_2$ on the right of $d_1$ \cite{GOF4} (the result, for $d_1,d_2$, will be denoted $[d_1|d_2]_D$ in $\diag$ and $[d_1|d_2]_L$ in $\ldiag$). These products endow $\diag$ and $\ldiag$ with the structure of monoids, with the empty diagram as neutral element. The corresponding commutative diagram is as follows (where $X^2$ means the cartesian square of the set $X$).

\begin{equation}
\begin{CD}
\textrm{Labelled diagrams}^2 @> Unlabelling^2 >> \textrm{Diagrams}^2 @> m_{(\L,\V)}\times m_{(\L,\V)} >> \textrm{Monomials}^2\\
@ V\textrm{product}VV				 @ V\textrm{product}VV		@ V\textrm{product}VV\\
\textrm{Labelled diagrams} @> Unlabelling >> \textrm{Diagrams} @> m_{(\L,\V)} >> \textrm{Monomials}\\
\end{CD}
\end{equation}
 
It is easy to see  that the labelled diagram (resp. diagrams) form free monoids. We denote  by  $\DIAG$ and $\LDIAG$ the $K$-algebras of these monoids \cite{GOF4} ($K$ is a field).\\
One can shuffle the product in $\ldiag$, counting crossings and superpositions. The definition of the deformed product is expressed by the diagrammatic formula 

\begin{center}
\includegraphics[scale=0.55]{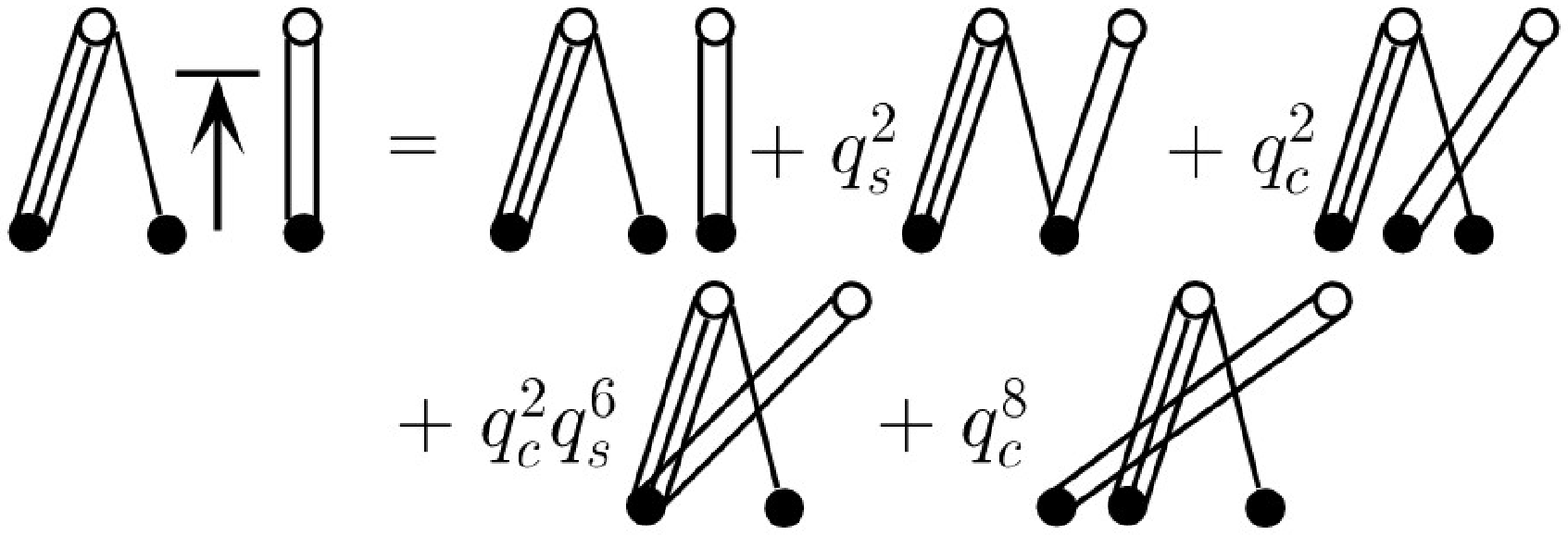}
\end{center}

\vspace{-2cm}
and the descriptive formula below.

\begin{equation}
	[d_1|d_2]_{L(q_c,q_s)}=\sum_{cs([d_1|d_2]_{L})\ are\ all\ crossing\ and\atop superpositions\ of\ black\ spots} 
	q_c^{nc\times weight_2}q_s^{weight_1\times weight_2} cs([d_1|d_2]_{L})
\end{equation}
where 

\begin{itemize}
    \item $q_c,q_s$ are coefficients in $K$
    \item the exponent of $q_c^{nc\times weight_2}$ is the number of crossings of ``what crosses'' times its weight
    \item the exponent of $q_s^{weight_1\times weight_2}$ is the product of the weights of ``what is overlapped''  
    \item terms $cs([d_1|d_2]_{L})$ are the diagrams obtained from $[d_1|d_2]_L$ by the process of crossing and superposing the black spots of $d_2$ on those of $d_1$, the order and identity of the black spots of $d_1$ (resp. $d_2$) being preserved.
\end{itemize}

What is striking is that this law (denoted above $\bar\ua$) is associative. Moreover, it can be shown \cite{DKPTT,GOF0} that this process decomposes into  two transformations: twisting and shifting. In fact, specialized to certain parameters, this law is reminiscent of others \cite{Ca}.

\begin{center}
\begin{tabular}{c|| c|c| c}
Parameters 	& $(0,0)$ (shifted)	& $(1,1)$ (shifted) & $(1,1)$ (unshifted)\\
\hline
Laws 				& $\LDIAG$  				& $\MQS$  					& Hoffmann \& Euler-Zagier 
\end{tabular}
\end{center}

\section{Hopf Deformation}

Using a total  order on the monomials of $\ldiag$, it can be shown that the algebra $\LDIAG(q_c,q_s)$ is free. Thus one may construct a coproduct $\Delta_t;\ t\in K$ such that 
$\LDIAG (q_c,q_s,t)=\left(\LDIAG (q_c,q_s),1_\ldiag,\Delta_t,\ep,S\right)$ is a Hopf algebra. We have the following specializations

\begin{center}
\begin{tabular}{c||c|c}
$(q_c,q_s,t)=$ 	& $(0,0,0)$ & $(1,1,1)$\\
\hline
$\simeq$ 				& $\LDIAG$  & $\MQS$  					 
\end{tabular}
\end{center}   

\section{Conclusion}

The results which we announced here in this note can be illustrated by the following picture. All details will be given in forthcoming papers \cite{DKPTT,GOF0}.

\begin{center}
\includegraphics[scale=0.5]{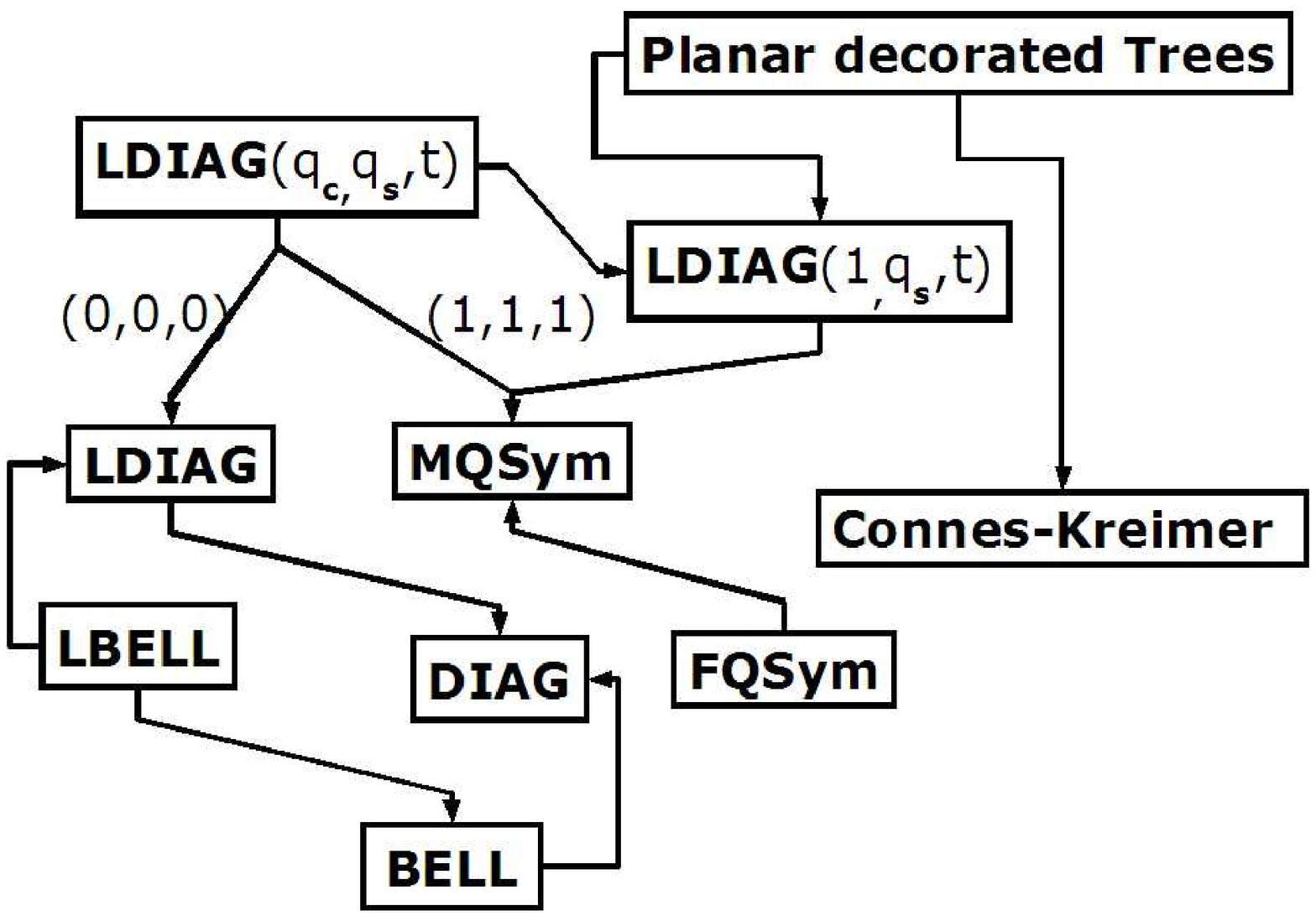}
\end{center}

\end{document}